# An improved method of delta summation for faster current value selection across filtered subsets of interval and temporal relational data

Derek Colley
School of Digital, Technologies and Arts
Office S343, Staffordshire University
Stoke-on-Trent, ST4 2DE, United Kingdom
derek.colley@staffs.ac.uk

Md. Asaduzzaman
School of Digital, Technologies & Arts
Office B009A, Staffordshire University
Stoke-on-Trent, ST4 2DE, United Kingdom
md.asaduzzaman@staffs.ac.uk

*Abstract*—Aggregation in relational databases is accomplished through hashing and sorting interval data, which is computationally expensive and scales poorly as the data volumes grow. In this paper, we show how quantitative interval and time-series data in relational attributes can be represented using delta summary values rather than absolute values. The need for sorting to determine the row corresponding to some maximum timestamp is negated, reducing the time complexity from at least O(n log(n)) towards O(n) and improving query execution times. We illustrate this new method in the relational algebra, present the implementation algorithmically, and test an implementation in two leading RDBMS products against the use of normal equivalents. We found this delta summation technique to be most effective for use cases with additive, numerical data upon which it is necessary to frequently obtain the latest values, and where the row cardinalities are in the order of $10^5$. Our experiments found the proposed new delta summation technique could execute faster than the equivalent standard selection method by up to 22.4%, reducing the overall query cost in some circumstances by up to 24.0%, reducing I/O load by up to 60.6%, but with increased query costs for write operations, an increase in CPU time and memory allocation, uncertain performance with very low or very high cardinalities and inconsistent results across different RDBMS platforms.

## I. Introduction

The relational model and SQL standards are notably poor at accommodating ordinal or temporal data, having little support for interval calculations. We define ordinal data in the usual sense, that from which distinct values relate to one another from some ordered domain and are separated by some interval, and temporal data here as a subclass of ordinals, as an instance of a tuple in a relation that includes a date or time component, and for which there are multiple tuples relating to the instance, and for which the instance is defined by some class attribute. For example, capturing a sensor reading $s$ at a regular interval will result in recording one tuple of data per period $p$, and over some total time $t$ there will be $(|p|+1)$ instances of $s$. Other sensor readings will take a different identifier but may be co-located in the same table. The user who wishes to know the latest sensor reading for some particular class of $s$ in the dataset is hard-pressed to write a well-performing query against the database since the data to request must be hashed into buckets, where all other involved attributes are grouped by the class attribute and the latest reading is identified by a MAX() aggregation over the date/time attribute which requires sorting the data. This is what we term the absolute summation method.

While any performance issues could be mitigated by strategies such as partitioning or indexing, the method itself is inherently inefficient since the search is a sort over some partitioned (or predicated) subset with the MAX() taken as the highest value of the sorted list, or the SUM() as the sum of the list. When millions or billions of rows are present, this sort is not scalable with a time complexity of at least the sort cost, where n is the cardinality of the table and the sort cost can vary depending on technique; bubble sort has O(n$^2$) complexity, merge sort has at best O(n log(n)) complexity and so forth. This compares favourably to our proposed method of delta summation which eliminates the sort.

Fig. 1 illustrates the different approaches between the absolute and the delta summation methods. Note how the absolute method (left) requires hashing the input rows and sorting each hash bucket, but the delta method (right) requires only a summation and a single scalar conversion, meaning the relative cost of running this query may be lower.

We tested the delta method of data insertion and summation to tables with record insertion which replaced date-time attribute values with delta calculations against all other values for the attribute (with common identifying keys), and we will show how alternative queries can be used to retrieve data from tables updated by this method faster than from tables with absolute values.

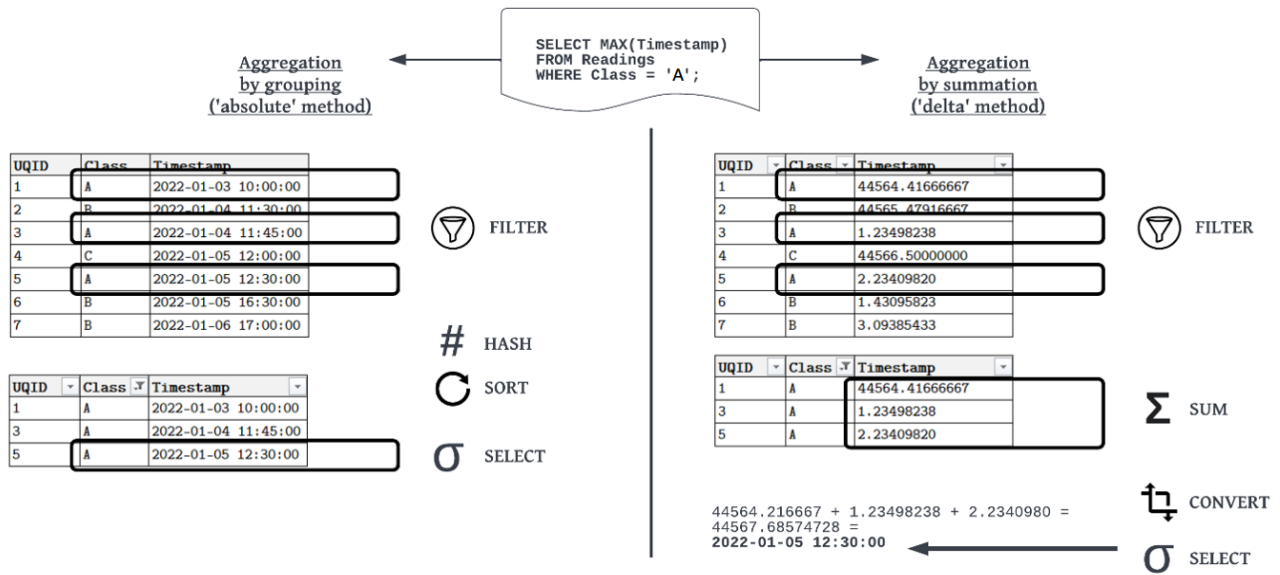

Fig. 1: Comparing methods for single-column aggregation: absolute vs. delta approaches

## II. Related Work

Relational databases are constructs based on the relational model proposed by Codd [1], later extended and developed by Date and Warden [2] and many others, and which has culminated in a mature, well-understood data persistence paradigm widely employed today. According to a popular benchmark measure, 70% of databases in current use are based on the relational model [3]. Relations (or to use Date's terminology, relational variables [2]) are set-theoretic constructs defined as a set of data held in one or more entities (relations, or tables), within which the values are arranged as tuples, or rows, each tuple element corresponding to an attribute, or column, and where columns inherit a datatype from some predefined domain type. Relations can be augmented by features such as constraints, views and indexes, and the relational model allows for complex relationships to be defined between entities forming a web of relational integrity within the database corresponding to some set of rules such as third normal form, or alternatively a data warehousing schema [4][5]. The maturity and ubiquity of the relational model means there are many millions of software implementations dependent on relational database platforms.

The relational model of databases is noted for being inflexible, pre-defined and intolerant to non-set theoretic operations, leading to misalignment with modern object-oriented paradigms, a phenomenon known as object-relational impedance mismatch [6][7][8][9]. Being set-theoretic constructs, one aspect of this mismatch manifests as a lack of native support for aggregation functions (such as MIN(), MAX(), AVERAGE()) within the underlying pure set theory, since there are no specific axioms under the Zermelo-Fraenkel (ZFC) standard that deal with aggregation [10]. Instead, classical computer science approaches, often based on object-oriented algorithms, are used as supplements. For example, in the Microsoft SQL Server platform, aggregation by MAX() is accomplished through reading all input rows, classifying each row into hash buckets, sorting each hash bucket, applying predicates (filters) and extracting the maximum value from the hash buckets where the predicates are true [11]. This is a programmer's approach to the problem and implies at least $O(n)$ time complexity for the table scan plus overhead for the hashing operation, the sort and the extraction. Such approaches have been jocularly termed 'RBAR', or row-by-agonising-row [12].

The cost, expressed as time complexity, for different hashing algorithms varies; Dietzfelbinger et al. [13] contend that uniform hashing given random input data results in $O(\log(n)/\log \log(n))$ complexity. Database vendors are silent on the hash algorithms used in aggregation, being largely proprietary information; however, some clues are found in older literature, particularly Aoki [14] who states that PostgreSQL uses (or used, in 1991), linear hashing algorithms; citing Litwin [15], who concludes the linear hashing process to have approximately $O(1.3n)$ time complexity, although his results vary from $O(n)$ to $O(1.77n)$ for successful searches (i.e. records found according to some predicate). Newer hashing algorithms claim time complexity $O(1)$, for example the Dynamic Perfect Hashing algorithm [13], or Cuckoo Hashing [16].

While it is reasonable to assume that modern database technologies use hashing algorithms operating at a circa $O(1)$ average, the same cannot be said of the time complexity of sorting algorithms. At best, algorithms like quicksort operate in $O(n \log_2(n))$, at worst $O(n^2)$, and this represents one of the better sorting algorithms currently known to computer science [17]. This means a database aggregation process dependent on hashing and sorting input to produce some aggregated result set is facing an immediate disadvantage against a method that does not involve sorting, particularly if such a method has a lower time complexity than the hash and sort operations combined, and a disadvantage that scales at least linearly and at worst exponentially as the volume of inbound data grows.

In so-called 'big data' environments, this scalability concern is significant. Partitioning is a useful method of overcoming high row cardinality by reducing the search space, where some partition key is selected, and the relation is separated into horizontal non-overlapping groups of tuples [18]. Consequent hash and sort operations then act upon a reduced tuple count when partitions are used, necessitating fewer row reads and the complexity therefore reduces to $O(n/p)$ where $p$ is the count of partitions, assuming all partition tuple membership cardinalities are equal – if unequal, then the time complexity still remains lower than O(n), providing $p >= 2$ and where the total count of partitions in use for the query is lower than the total $p$. This holds for hash and sort operators too, where the divisor $p$ or derivative thereof can be applied to whichever hash or sort algorithm is in operation. Partitioning also provides the advantage of easy parallelisation, where operations on disjoint problem spaces can be divided amongst different processor cores, facilitating concurrent calculation; although not specifically aimed at database architecture, Ye et al. [19] present three partition-and-aggregate algorithms which illustrate the performance gain in terms of aggregation over numerical vectors and the principles of which are applicable more generally to database aggregation operations.

Another technique to reduce the search space is the use of indexes to avoid reads of the whole relation. Relations without any indexes at all are heaps, and consequently reading all values from the heap means no guarantee of input-output page contiguity (although this may occur by accident and will decrease as the heap fractures due to page inserts, splits and reallocations). All RDBMSs offer some variety of indexing strategies, albeit with different terminology; B-tree indexes are universal, but PostgreSQL, for example, offers GiST indexes, which are hybrid indexes well-suited to queries using multi-dimensional or non-primitive operators [20]. Microsoft SQL Server uses hash indexes in memory-optimised tables [21] and it is becoming common to see columnar indexing, where values are indexed by column rather than by row, across most major platforms [22]. While indexes are useful for significantly reducing the search space of queries with narrow predicates, they are less useful when a significant portion of rows are required to be read to satisfy a query; this threshold is used by the Microsoft SQL Server query optimiser to determine whether a table or index read operation should be used, and the type of operation: scan or seek [23]. The inherent problem of facilitating faster aggregation exists with or without indexes, to a varying degree.

There is some evidence of recent literature from researchers seeking to improve aggregation methods in relational databases. Khamis et al. [24] examined better methods of aggregation for queries with functional additive inequalities in join conditions and presented a new approach. Schleich et al. [25] decomposed multi-relation query join trees and applied aggregate decomposition with parallelisation in order to solve aggregation problems in Online Analytical Processing (OLAP) environments, albeit in the abstract, rather than through demonstrable application. Böhlen et al. [26] present an overview of temporal table support in the SQL:2011 standard [27] and describe a new technique of temporal normalisation for faster executions of queries against temporal relations; relevant since temporal data is interval data, and applicable here. Cai et al. [28] provide a detailed taxonomical survey of aggregation approaches across relational and non-relational applications, including online aggregation [29], a process of splitting a large batch aggregate query into sub-aggregations where the progress of aggregation calculation is visible by the user. It is unclear if this approach is supported today by modern RDBMSs, although limited real-time query feedback is given in Microsoft SQL Server, for example, with Live Query Statistics [30] which includes progress through queries containing aggregate hashing operations.

To summarise, the problem of conducting efficient aggregations over interval data in relational databases remains current. There is little evidence of recent improvement in the base algorithms which perform aggregation across large query sets, inferring a significant research gap. It is this research gap that we hope to address through the presentation of delta summation as an alternative approach to traditional query aggregation in relational database management systems.

### III. METHODOLOGY

The methods presented here test whether summing across an attribute could be faster than multiple table or index scans across the relation incorporating unnecessary hashing and sorting, and so modifying the methods in which intervals or date/time values are stored from absolute to relative delta-aware values could yield a simpler and more performant selection query on read. We take a pragmatic approach, setting out the theory below, and testing this using positivistic empiricism through controlled comparative experimental analysis between the absolute (grouping) methods and the new delta methods we propose.

#### A. Relational Algebraic Representation

We first define our method in formal notation. We use a variation of Codd's relational algebra [1] extended to include assignments and aggregations. The pre-requisites are:

- Given a relation R comprising of a non-zero number of attributes ($R.a_1, R.a_2 \ldots R.a_n$), of which one attribute, denoted $R.a_i$, is of a timestamp type (or a numeric base type that maps to a timestamp type), forming a total time period $t$ comprised of $p$ discrete intervals as a domain from which $R.a_i$ forms a subset thereof;

- And given the relation R includes at least one identifying non-primary key column (class attribute) that identifies a subset of tuples within the relation, which we denote $R.a_x$;

- And given the object is to determine the latest (maximum) timestamp within $R.a_i$ for some $R.a_x$, subject to one or more optional predicates ($\varphi_0 \ldots \varphi_n$), or the summation of the interval data:

Then the existing absolute method of grouping in a subquery producing output set S can be expressed as a selection (1):

$$S = \sigma[R.a_x, (R.a_1 \ldots R.a_n)] \text{ AS X}$$
$$\theta \; \varphi[R.a_x \; \Upsilon \; \text{MAX}(R.a_i) \,|\, \text{SUM}(R.a_i) \,|\, \text{AS } R.a_i] \text{ AS Y}$$
$$(X.R.a_x = Y.R.a_x \text{ AND } X.R.A_i = Y.R.a_i)$$

Given a set of 0 or more predicates ($\varphi_0 \ldots \varphi_n$) on $R$.

(1)

Expression (2) shows an insertion of a value $c$ into $R.a_i$ subject to some predicate $R.a_x$ using the ordinary, or absolute, method.

$$R.a_i \leftarrow [c, R.a_x]$$

(2)

We propose the replacement of R.a$_i$, the interval attribute, with the delta, or difference, between the preceding attribute value in the manner that a tuple would ordinarily be collected. Thus, on tuple insertion, we replace the current value $c$ being inserted into R.a$_i$ on insertion with a delta value of $c$ against the sum of all previous values of R.a$_i$, with R.a$_x$ used to group the value as previously described. This moves the computational load from the aggregate selection to the insertion operation.

The selection operation on this alternative representation is shown in (3) and this can be used for either the MAX() and SUM() aggregations which are functionally identical when calculating deltas:

$$S = \sigma[R.a_x, \text{SUM}(R.a_i) \text{ AS } R.a_i] \quad (3)$$

The insertion operation on the alternative representation is shown in (4):

$$R.a_i \leftarrow [\text{iff. } R.a_i == \varnothing : R.a_i = c,$$
$$\text{else } R.a_i = (c - \sum[R.a_i - 1]) \; \varphi \; (R.a_x)] \quad (4)$$

Consequently, the selection operation S is simplified from the absolute method, using aggregation and grouping (1) to the delta method (3) using the SUM() function. Both (1) and (3) are functionally identical, if syntactically different, but we theorise Expression (4) to be less complex at runtime and consequently exhibit faster performance characteristics during execution than (2).

We contend that although the insertion operation in (4) necessary for the alternative selection operation is more computationally complex than the absolute method of insertion in (2) and therefore may incur a higher execution cost (in terms of query execution time and resources consumed), there may also be a significant saving in execution cost in the same terms between the inefficient form of the query selection operation in (1) and the simplified form presented in (3). It is noteworthy that insertion operations on data like these are 'blind', in the sense that there is likely to be no application or person on the other end of the process waiting for a result, where selection operations are often useful in real-time. Delays, therefore, in insertion are arguably more palatable than delays in selection. Successful completion of insertion operations will depend, of course, on the completion and committal of the transaction as per all other database transactions.

## B. Algorithmic Representation

Algorithm 1 shows the general algorithm for the new selection method.

ALGORITHM 1: RELATIONAL SELECTION OF AN ORDINAL NUMERIC ATTRIBUTE STORED USING DELTA SUMMATION

| |
|---|
| **Given:** $R$ as a relation (alternatively a table), comprised of attributes ($R.a_1 \ldots R.a_n$) |
| That $R$ contains one ordinal nominal value on which to aggregate, denoted $R.a_i$ |
| That there exists a candidate key column R.a$_x$ by which R.a$_i$ can be grouped |
| That $c$ represents a value and member of R.a$_x$ used to group by in the selection query |
| That Q represents the selection query containing $c$ and a group by operation on $R.a_i$: |
| 1. For Q: |
| 2. → Replace Q with Q', where Q' is defined as: |
| 3. → → Select the sum of all values in the attribute R.a$_i$ grouped by R.$a_x$ |
| 4. → → (Optional) Filter the output of Step 3 by some predicate(s) |
| 5. Execute Q'. |
| **End** |

Algorithm 2 shows the general algorithm for the new insertion method.

ALGORITHM 2: RELATIONAL INSERTION OF AN ORDINAL NUMERIC ATTRIBUTE STORED USING DELTA SUMMATION

| |
|---|
| **Given:** $R$ as a relation (alternatively a table), comprised of attributes ($R.a_1 \ldots R.a_n$) |
| $C$ as a set of 1 or more ordinal numerics, such as timestamps, for insertion into $R.a_i$ |
| This algorithm only fires on receipt of an INSERT query $Q$ on $R$ containing some $C$ |
| That there exists a candidate key column R.a$_x$ by which R.a$_i$ can be grouped: |
| → And that the value $C$(k) represents the candidate key for some values of $C$. |
| 1. For Q: |
| 2. → Replace Q with Q', where Q' is defined as: |
| 3. → → For each $c$ in $C$: if the cardinality of $R$ on $R.a_i = c$ is the empty set: |
| 4. → → → Insert the value $c$ to R.a$_i$ |
| 5. → → Else: |
| 6. → → → Calculate the sum of all $R.a_i$ where $R.a_x = C$(k), assign to variable $X$ |
| 7. → → → Calculate ($c - X$) and insert into $R.a_i$ instead of $c$ |
| 8. Execute Q'. |
| **End** |

## IV. IMPLEMENTATION AND TESTING

We tested insertion and selection operations against both the standard methods and our proposed delta summation methods, the selections with and without an additional predicate, in two Relational Database Management Systems (RDBMSs), Microsoft SQL Server 2019 (Developer Edition) running on Windows 10 on an Intel Core i7 processor (single socket, 4-core 1.3GHz/1.5GHz with 16GB memory, the service capped to 12GB, on solid-state storage) and PostgreSQL 13.4 running as an RDS instance on 64-bit RedHat Linux hosted by Amazon Web Services, on the db.t2.small RDS instance class (Intel AVX, 1 core 2.5GHz, 2 vCPU, 2GB memory) on solid-state storage. Network transmission bandwidth and latency are irrelevant since the test queries are identical (barring minor syntactic differences) between the platforms and the return dataset sizes are negligible. We discard wait time for result set transmission from the experimental results accordingly.

Two platforms, 3 sets of row cardinalities, selection and insertion testing, with and without predicates (for selection; insertions do not take predicates) formed 18 sets of experiments in total, each of which were repeated in 10 iterations over a short period to obtain 180 sets of readings. Each set of readings was comprised of 5 metrics, yielding 900 data points; the whole test set was repeated for the delta data set, yielding 1,800 data points in total.

To further prevent the influence of confounding factors, for Microsoft SQL Server we cleared the buffer and query caches each time to force homogeneous parse and compile operation impacts across each experiment; for PostgreSQL, no such commands exist, and so we restarted the service between tests to achieve the same effect. In the Results section of this document, we normalise our findings across both RDBMSs when comparing across platforms to eliminate the performance characteristics of each platform as a cause factor.

For all experiments we used a relational table comprising of a primary key column (PKC), a non-primary key classification column (CC) and an interval column as a timestamp (IC). We seeded the table with 10 classes in the CC column and 100,000, 1,000,000 and 10,000,000 randomly generated interval values (three cardinality sizes) in a 1-year range at a granularity of 1 microsecond in the IC column. All classes in the CC and PC columns were randomly distributed. We used the same table data, once generated, for the experiments on the standard methods of selection and insertion.

For the delta summation methods, we transformed the delta summation table by recalculating the values in the IC column as running interval totals, partitioned by the CC column and ordered by the PKC column. We derived the durations from the *NIX epoch date representations, accurate to the microsecond. We then tested selection and insertion using implementations of our algorithms. We tested with and without a predicate on the CC column. For Microsoft SQL Server testing, we recreated the tables exactly as in PostgreSQL including the primary key columns, imported the data directly from PostgreSQL using SQL Server Integration Services and conducted sampling tests to ensure both data sets were identical. The supplementary material, code and results data for both the PostgreSQL and MSSQL configurations are provided on Github [31].

To measure relative performance, we used the following metrics on both platforms:

TABLE 1: MEASUREMENT METRICS

| Metric | Range | Type | Comment |
| --- | --- | --- | --- |
| CPU time (ms) | 0-∞ | Real | From execution plan |
| I/O (Reads) | 0-∞ | KB | Page reads, converted to KB (8KB pages) |
| Memory Grant | 0-∞ | KB | Memory granted to query |
| Total Subtree Cost | 0-∞ | Real | Relative cost from execution plan |
| Elapsed Time (ms) | 0-∞ | Real | Duration in milliseconds |

To reduce the complexity of the experiments and control the number of independent variables, we exclude the use of indexing and partitioning from scope. For measurements, in MS SQL we used the *sys.dm_exec_query_stats* dynamic management view; in PostgreSQL, we used the *pg_stat_statements* view, which entailed loading the *pg_stat_statements* to the *shared_preload_libraries* of the instance which we did via a custom parameter group in AWS RDS for the database instance. It is not possible to flush caches manually in PostgreSQL so we instead controlled for query plan re-use through labelling each query uniquely, forcing recompilation, and restarting the service after each complete batch of tests. We achieved this using a test harness written in Python which could remotely restart the RDS via the AWS CLI and communicate with PostgreSQL using the psycopg2 package.

## V. RESULTS

### A. Microsoft SQL Server

We start by examining the descriptive statistics for the control sets vs. the delta summation sets.

The uniformity of the measurements of elapsed time is evident when we examine the data by iteration, rather than in the aggregate. Fig. 2 shows each type of test, arranged by iteration on the x-axis. There is a clear similarity between iterations which is borne out by the analysis of the deviations in Table 2. For the 100k row tests, we note that the delta summation method outperformed the control method (in terms of elapsed time) by approximately 1.4%, but the control method outperformed the delta summation method when querying without predicates and for the insertion by up to 200%. This may be explained by the relatively low row counts, since for a query without predicates the hash is unnecessary, and the sort operation outperformed the addition operation; likewise for the insertion, we hypothesised insertion as a more expensive operation for delta summation which is shown by our results.

The abbreviations in the following figures and tables are interpreted as follows: Control (C)/Delta Method (D) | With (W)/Without(WO) Predicates (P) | Insertion (I).

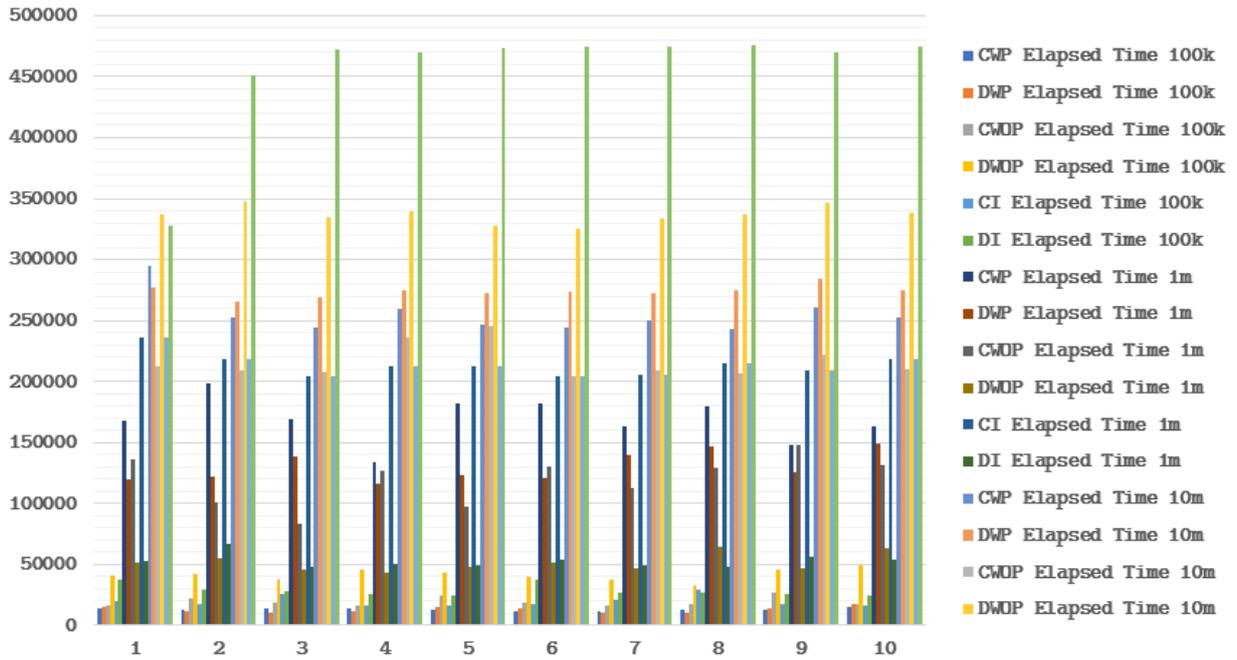

*Fig. 2: Comparison of elapsed time (μs) across iterations for all test types*

The trends are reversed for higher row counts. we note the delta summation method outperforming the control method for every type of test, consistently so with low (sub-5%) standard deviation across iterations. For 1m rows with predicates, the delta summation method had an average 22.7% lower execution time than its counterpart. This even held true without predication since the time complexity advantages of addition over sort become apparent. The surprising result here was the longer execution time for the control insertion rather than the delta method insertion; the cause is unknown, but the phenomenon is repeated over all iterations.

For the largest row counts (10m), the data shows the delta summation method is inferior to the control method. We note at this scale there is also far more unpredictability in the results between iterations, with up to +/- 44.9% variation. For selection with predicates, there is a 6.9% execution time advantage to the control method. With the deviations so high between iterations it is possible some confounding variables were at play; for example background compute activity on the platform, or inconsistent I/O speeds. However, if we take the results at face value, the delta summation method appears to start to break down at extremely high table cardinalities; the difference in the select with predicates tests in particular is not large, but significant.

As shown in Table 2, we noted very low standard deviation for each test series from the mean: typically +/- 0.25% – 0.93% for all tests with 100k rows; worsening to +/- 1.07% - 4.39% for all tests with 1m rows and the test series with 10m rows proving most unstable at +/- 5.39% - 44.93% deviation. This would generally indicate the performance of most tests is repeatable and predictable, an important property of any eventual solution, with degradation shown at very high row counts.

Examining I/O activity, we noted a significant decrease in the reads required from memory and the underlying file system. Table 3 describes a cost reduction for every single test run where delta summation is shown to require fewer I/O operations than the control case. For 1m rows with predicates, the reduction in reads required was 60.59%, which is as significant as the reduction in elapsed time in the same case.

TABLE 2: DESCRIPTIVE STATISTICS FOR ELAPSED TIME, FOR ALL TESTS ACROSS ALL ITERATIONS

| Test Type | Range | SD | Mean | Median | SD +/- Mean % | SD +/- Median % |
|---|---|---|---|---|---|---|
| CWP Elapsed Time 100k | 4350 | 1166.76 | 199374.11 | 235940.00 | 0.29 | 0.25 |
| CWOP Elapsed Time 100k | 7873 | 2431.67 | 214585.24 | 217943.00 | 0.57 | 0.56 |
| CI Elapsed Time 100k | 10569 | 3530.04 | 214659.58 | 208189.00 | 0.82 | 0.85 |
| DWP Elapsed Time 100k | 15833 | 4365.85 | 221797.12 | 235754.00 | 0.98 | 0.93 |
| DWOP Elapsed Time 100k | 12794 | 4248.74 | 221101.96 | 245725.00 | 0.96 | 0.86 |
| DI Elapsed Time 100k | 12926 | 1166.76 | 213784.54 | 204892.00 | 0.27 | 0.28 |
| CWP Elapsed Time 1m | 64299 | 4473.54 | 223293.79 | 208908.00 | 1 | 1.07 |
| CWOP Elapsed Time 1m | 32014 | 17434.48 | 222819.21 | 215122.00 | 3.91 | 4.05 |
| CI Elapsed Time 1m | 64551 | 11412.52 | 229583.65 | 221924.00 | 2.49 | 2.57 |
| DWP Elapsed Time 1m | 21955 | 19133.53 | 222579.08 | 218063.00 | 4.3 | 4.39 |
| DWOP Elapsed Time 1m | 31471 | 7065.64 | 219341.23 | 216441.40 | 1.61 | 1.63 |
| DI Elapsed Time 1m | 18920 | 8958.64 | 218989.31 | 212151.00 | 2.05 | 2.11 |
| CWP Elapsed Time 10m | 52499 | 5402.41 | 28950.71 | 28950.71 | 9.33 | 9.33 |
| CWOP Elapsed Time 10m | 17975 | 14662.88 | 16318.94 | 16318.94 | 44.93 | 44.93 |
| CI Elapsed Time 10m | 40833 | 4479.34 | 22656.17 | 22656.17 | 9.89 | 9.89 |
| DWP Elapsed Time 10m | 22116 | 13082.73 | 17599.36 | 17599.36 | 37.17 | 37.17 |
| DWOP Elapsed Time 10m | 31471 | 6681.27 | 19076.14 | 19076.14 | 17.51 | 17.51 |
| DI Elapsed Time 10m | 147568 | 8958.64 | 78263.32 | 78263.32 | 5.72 | 5.72 |

TABLE 3: COMPARISON OF I/O OPERATIONS FOR THE CONTROL AND DELTA CASES

| Test Type | Row Count | Control | | | Delta | | | % Diff Total Reads |
|---|---|---|---|---|---|---|---|---|
| | | Logical Reads | Physical Reads | Total Reads | Logical Reads | Physical Reads | Total Reads | |
| With Predicates | 100000 | 455 | 922 | 1377 | 330 | 684 | 1014 | -26.36% |
| Without Predicates | 100000 | 455 | 922 | 1377 | 330 | 684 | 1014 | -26.36% |
| Insertion | 100000 | 456 | 922 | 1378 | 330 | 680 | 1010 | -26.71% |
| With Predicates | 1000000 | 4511 | 9116 | 13627 | 3262 | 2108 | 5370 | -60.59% |
| Without Predicates | 1000000 | 4511 | 9116 | 13627 | 3262 | 6556 | 9818 | -27.95% |
| Insertion | 1000000 | 4512 | 9102 | 13614 | 3262 | 6556 | 9818 | -27.88% |
| With Predicates | 10000000 | 32974 | 65942 | 98916 | 32564 | 52551 | 85115 | -13.95% |
| Without Predicates | 10000000 | 32974 | 65942 | 98916 | 32564 | 65230 | 97794 | -1.13% |
| Insertion | 10000000 | 32975 | 65942 | 98917 | 32564 | 65230 | 97794 | -1.14% |

We noted zero memory grant allocated for the tests with the lower row counts; this is due to sufficient memory already allocated in the various pools to service the query. However, as the row count grew, we observed memory grants steadily increasing. Table 4 provides an overview.

TABLE 4: COMPARISON OF MEMORY GRANT ALLOCATIONS (USED GRANT KB) BETWEEN CONTROL AND DELTA CASES

| Test Type | Control Used Grant KB | Delta Used Grant KB | Percent. Diff. |
|---|---|---|---|
| 100k With Predicates | 0 | 0 | 0.00% |
| 100k Without Predicates | 0 | 336 | n/a |
| 100k Insertion | 0 | 0 | 0.00% |
| 1m With Predicates | 0 | 0 | 0.00% |
| 1m Without Predicates | 688 | 5639 | 719.62% |
| 1m Insertion | 688 | 3398 | 393.90% |
| 10m With Predicates | 2854 | 2849 | -0.18% |
| 10m Without Predicates | 2952 | 6009 | 103.56% |
| 10m Insertion | 2970 | 3200 | 7.74% |

In general, we note the memory grant required by the delta summation method to be higher in almost all cases than the control method, especially as row cardinalities increase. We speculate this is due to the requirement to keep all row values in memory during the addition, rather than discarding hash buckets that do not match predicates, but the reason is unclear.

Next, we examined the query cost via the query execution plans. The plans demonstrate the difference in execution methods between the control and delta summation techniques. Although it appears that the delta summation method has an extra step, in testing, the stream aggregate component (which calculates row aggregates across groups – the hash mechanism) takes more time and resources to execute with the control query, and as a result the total subtree cost for the control query is 4.97 vs. 3.75 for the delta summation query, a reduction in cost of approximately 25% using the new method.

The total subtree costs of the execution plans for each test are given in Table 5. We note for all selections, there was a marked decrease in total subtree cost (i.e. cost of the execution plan) but for all insertions, the cost was higher with the delta summation method than the control method, which we theorised would be the case.

TABLE 5: COMPARISON OF TOTAL SUBTREE COSTS BETWEEN ALL TEST TYPES FOR ALL ITERATIONS

| Test Type | Row Count | Control - Total Subtree Cost | Delta - Total Subtree Cost | % Difference |
|---|---|---|---|---|
| With Predicates | 100000 | 0.50 | 0.37 | -26.00% |
| Without Predicates | 100000 | 0.50 | 0.41 | -18.00% |
| Insertion | 100000 | 0.52 | 0.83 | 59.62% |
| With Predicates | 1000000 | 4.98 | 3.75 | -24.70% |
| Without Predicates | 1000000 | 4.50 | 4.11 | -8.67% |
| Insertion | 1000000 | 4.50 | 5.58 | 24.00% |
| With Predicates | 10000000 | 28.40 | 28.10 | -1.06% |
| Without Predicates | 10000000 | 27.30 | 27.00 | -1.10% |
| Insertion | 10000000 | 27.30 | 55.10 | 101.83% |

Finally, we examined the worker time (CPU cost) as a relative measure using the same methodology as for the elapsed time.

We note significantly increased CPU time as row counts increase; for all cases, CPU worker time was lower for the delta summation method than the control method where the row count was 100,000; else, the CPU worker time significantly increased as the row cardinalities increased. This proves an increased use of CPU resources for the delta summation method that scales exponentially; at odds with the lower overall execution time, meaning that despite more CPU time being required by the delta summation method, it is still capable of outperforming the control method in terms of execution time in many cases.

As with the elapsed time metric, we note fairly consistent performance across all iterations for all tests, indicating reliability. The results indicate high variation when working with large row counts between worker times as the predictability of CPU consumption starts to decay. As with the previous figures, CPU use is significantly higher for the delta summation method especially as row counts increase.

*B. PostgreSQL*

*Addendum on query collection*: It was necessary to get the EXPLAIN data values out from the console for the query cost and CPU cost; we modified the estimate_cost() function provided by Beldaz [32] to ease this process. This yielded the plan cost and rows, incorporated into the PostgreSQL findings. For CPU use, there is no native field available in the system views, but we were able to derive the cpu_tuple_cost from the overall plan cost by use of the guidance in the execution plan system documentation [20], particularly the formula:

query cost =
  (disk pages read * seq_page_cost)
    + (rows_scanned * cpu_tuple_cost)

which rearranged becomes:

cpu_tuple_cost =
  ( query cost / rows_scanned )
    – ( disk pages read * seq_page_cost )

We obtain query cost from the *estimate_cost()* function; rows scanned and disk pages read from the appropriate columns of the *pg_stat_statements* system view; and note that *seq_page_cost* defaults to 1.0, which is the setting in this test instance of PostgreSQL. We can then derive the SQL query to capture all metrics of interest for a given query, using the method demonstrated in Code Listing 1. There are no per-query memory statistics available in this platform, so these are excluded in our PostgreSQL results.

```
SELECT pg_stat_statements_reset();

-- TEST 1: PGSQL, 100K, SELECT WITH PREDICATES
SELECT  to_timestamp("S") FROM
(     SELECT SUM("DateTimeIncrement") AS "S"
    FROM "deltaTestRandom100k"
    WHERE "Category" = 'E') AS src

;WITH queryplandata AS (
SELECT totalcost, planrows
FROM estimate_cost (  '
SELECT  to_timestamp("S") FROM
(     SELECT SUM("DateTimeIncrement") AS "S"
    FROM "deltaTestRandom100k"
    WHERE "Category" = ''E'') AS src' ) )
,
runstats AS (
SELECT total_exec_time, shared_blks_read,
local_blks_read, blk_read_time, rows
FROM pg_stat_statements WHERE "query" like
('%to_timestamp%')
                        AND rows > 0 AND
query not like ('%estimate_cost%') )
,
cpustats AS (
select relpages, reltuples from pg_class where
relname = 'deltaTestRandom100k' )

select  runstats.total_exec_time,
runstats.shared_blks_read,
runstats.local_blks_read,
        runstats.blk_read_time, runstats.rows,
        cpustats.relpages, cpustats.reltuples,
        (queryplandata.totalcost /
runstats.rows) - (cpustats.relpages * 1.0) AS
cpu_cost_per_tuple_relative
from    runstats, cpustats, queryplandata
```

*Code Listing 1: General form for obtaining query metrics from PostgreSQL*

As with Microsoft SQL Server, we start by examining the query execution times. We observe the delta summation technique significantly outperforms the control technique for $10^5$ rows; it also outperforms the control technique at $10^4$ rows but falls behind the control at $10^6$ rows. Once again, we see negligible variation between iterations at low row cardinalities but up to +/- 99.6% variation at 10m rows, as observed for the other platform.

It is notable that the first iteration of tests shows data outliers far in excess of the mean, where the delta summation method with and without predicates for the first iteration only took much longer than expected. These readings appear to be anomalous and may correlate with some other environmental confounding factor since other readings taken in the same time period are also somewhat deviant from the expected values.

Examining the I/O measurements, we found a surprising result. The reads for the delta summation method in PostgreSQL were almost exactly twice that of the control method, despite an identical methodology, and the reverse found for Microsoft SQL Server, for all row cardinalities. The exactitude of the difference implies that the rows were read twice for the delta summation, and the difference may be explained by the differences in the methodologies of hash and sort operations between the platforms.

Unlike in Microsoft SQL Server, we were able to obtain the plan costs at runtime. We observe something interesting; the PostgreSQL engine shows no sign of hashing or sorting in the plan for the control query; instead, a filter is applied on the table scan ('*seq scan*') as with the control query, and a MAX() aggregate is applied. It is unclear from the plan whether hashing or sorting is taking place in the background or whether some other method or algorithm is being applied. In this example, the query plan cost is significantly lower for the control query than the delta summation query. This holds true for all cases with insertions in particular performing extremely badly against the control case.

We tested whether the number of reads was correlated with the total subtree costs. We found the correlation co-efficient to be 0.97 and an almost perfect correlation with the p-value = $5.83786 \times 10^{-14}$. This means the I/O cost is almost certainly responsible for the higher plan costs.

Finally, we examined the CPU worker time. We found that for all selections, the difference between the control and the delta summation test results are practically indistinguishable. For insertions, as theorised, the delta cost is higher. While this set of results may not add much significance to the overall findings, it is circumstantial evidence that the control and delta summation query sets are functionally identical, having incurred the same CPU load per tuple, which adds credence and confidence that our experiments are fundamentally sound.

## VI. SUMMARY OF FINDINGS

The test results from the Microsoft SQL Server platform were as follows. The delta summation method resulted in average lower execution times of 22.4% where row volumes were substantial (in the region of $10^5$ rows); but was outperformed by the control method for very small ($10^4$) and very large row counts ($10^6$+). The reliability of the results for very large row counts is questionable due to high variation (+/- 44.1%) between iterations, possibly due to environmental confounding factors. We noted a substantial decrease in I/O required for the delta summation method, ranging from -1.14% to -60.59%, meaning fewer disk accesses are required to run these queries; we also noted a reduction in query plan cost ranging from -1.1% to -24.0% for selection queries, although as predicted, insertion query costs rose significantly. Memory consumption increased for delta summation queries by up to 8x; likewise, CPU worker time for these queries was over twice that for the control queries in 7 out of the 9 test sets, although the reliability also suffered for very large row cardinalities (106), due to high variation between iterations (+/- 40.4%).

For PostgreSQL, the pattern of execution time savings was inconsistent. We noted faster execution (+10.6%) of the delta summation method without predicates for 105 rows (see Fig. 16) but much worse performance on insertions at all cardinalities and heavily-variable performance for all other test classes, tending towards better performance for the control queries. We noted also the high unexplained variance for 106 rows observed in Microsoft SQL Server. Unlike in Microsoft SQL Server, the PostgreSQL versions of these queries resulted in doubling I/O reads, which correlated almost perfectly (r = 0.97, p < 0.01) with the increase in query costs. We theorise that the cause is either an inferior MAX() aggregation method in PostgreSQL compared to Microsoft SQL Server, or a superior hash and sort function. Memory grant information was unavailable in this platform. We compared CPU worker time per tuple and found no appreciable difference in any of the test cases except insertion, implying good consistency between the operations in the control and delta summation techniques but no overall findings for this metric.

## VII. CONCLUSIONS AND FUTURE WORK

In this research, we sought to establish, for tables containing interval data columns, such as timestamps, and where some secondary attribute can function as a predicate, whether calculating, storing and retrieving the latest values from this table is more efficient than the current technique of hashing each row group by predicate value, sorting the groups and selecting the maximum. We showed that in theory, we can reduce the time complexity from O(n log(n)) to O(n) by eliminating the sort, although in practice the growth of the execution time was exponential, and we presented the relational algebra and associated algorithms.

After testing, for the first platform we found the delta summation method we proposed to run faster than the current method (> 22%), drive down the associated query costs and incur fewer I/O reads, albeit at the expense of higher CPU load and memory grants. For the second platform, we found generally worsening execution times with the exception of delta summation queries without predicates in the order of $10^5$ row cardinality, indicating this method may be of some limited use on this platform; we found increased query costs and associated I/O reads, and higher CPU costs per tuple.

In summary, we find our method to be useful under certain circumstances; particularly where interval data is used with or without predication, against the Microsoft SQL Server platform more so than PostgreSQL and where the row cardinalities are in the area of $10^6$.

Future work in this area includes testing more RDBMS platforms, refining the delta summation algorithms and incorporating the use of indexes, materialised views and horizontal partitioning for maximum effectiveness when searching large tables.